\documentclass[preprint]{aastex6}

\begin{document}                                                                     

\title{The Components of Cepheid Systems: The FN Vel System
  \footnote{Based on observations with the NASA/ESA {\it Hubble Space Telescope }
    obtained at the Space Telescope Science Institute, which is operated by the
    Association of Universities for Research in Astronomy, Inc. under NASA
    contract NAS5-26555}
} 


\author{Nancy Remage Evans}
\affil{Smithsonian Astrophysical Observatory,
MS 4, 60 Garden St., Cambridge, MA 02138; nevans@cfa.harvard.edu}

\author{Pierre Kervella}
 \affil{LESIA, Observatoire de Paris, Universit\'e PSL, CNRS, Sorbonne Universit\'e, Universit\'e de Paris, 5 Place Jules Janssen, 92195 Meudon, France}

\author{Joanna Kuraszkiewicz}
\affil{Smithsonian Astrophysical Observatory,
MS 67, 60 Garden St., Cambridge, MA 02138; jkuraszkiewicz@cfa.harvard.edu}

\author{H. Moritz G\"unther}
\affil{Massachusetts Institute of Technology, Kavli Institute for Astrophysics and
Space Research, 77 Massachusetts Ave, NE83-569, Cambridge MA 02139, USA}

\author{Richard I. Anderson}
 \affil{Institute of Physics, \'Ecole Polytechnique F\'ed\'erale de Lausanne (EPFL), Observatoire de Sauverny, Chemin Pegasi 51B, 1290 Versoix, Switzerland}
 
\author{Charles Proffitt}
\affiliation{Space Telescope Science Institute, 3700 San Martin Drive, Baltimore, MD 21218}

\author{Alexandre Gallenne}
\affiliation{Instituto de Astrof\'isica, Departamento de Ciencias F\'isicas, Facultad de Ciencias Exactas, Universidad Andr\'es Bello, Fern\'andez Concha 700, Las Condes, Santiago, Chile and French-Chilean Laboratory for Astronomy, IRL 3386, CNRS, Casilla 36-D, Santiago, Chile}

\author{Antoine M\'erand}
 \affil{European Southern Observatory, Karl-Schwarzschild-Str. 2, 85748 Garching, Germany}

\author{Boris Trahin}
 \affil{LESIA, Observatoire de Paris, Universit\'e PSL, CNRS, Sorbonne Universit\'e, Universit\'e de Paris, 5 Place Jules Janssen, 92195 Meudon, France}

\author{Giordano Viviani}   
\affil{Institute of Physics, \'Ecole Polytechnique F\'ed\'erale de Lausanne (EPFL), Observatoire de Sauverny, Chemin Pegasi 51B, 1290 Versoix, Switzerland}

\author{Shreeya Shetye}   
\affil{Institute of Physics, \'Ecole Polytechnique F\'ed\'erale de Lausanne (EPFL), Observatoire de Sauverny, Chemin Pegasi 51B, 1290 Versoix, Switzerland
and Instituut
voor Sterrenkunde, KU Leuven, Celestijnenlaan 200D bus 2401, Leuven, 3001,
Belgium}


 \begin{abstract}
   Cepheid masses continue to be important tests of evolutionary tracks for
   intermediate mass stars as well as important predictors of their future fate.
   For systems
   where the secondary is a B star, {\it Hubble Space Telescope} ultraviolet
   spectra have been obtained.  From these spectra a temperature can be derived,
   and from this a mass of the companion M$_2$. Once {\it Gaia} DR4 is available, 
   proper motions can be used to determine the inclination of the orbit. 
   Combining mass of the companion, M$_2$,  the mass
   function from the ground-based orbit of the Cepheid and the inclination 
   produces the mass of the
   Cepheid, M$_1$.  The Cepheid system FN Vel is used here to demonstrate this
   approach and what limits can be put on the Cepheid mass for 
    inclination between 50 and 130$^o$. 
   
\end{abstract}




\section{Introduction}





More than a century after the discovery of the Cepheid Leavitt (Period--Luminosity) Law, 
Cepheids continue to be the first step in the extragalactic distance scale (Riess, et al 
2022; Breuval et al. 2022).  This is of particular importance because of the tension between 
the Hubble constant from Cepheids and Type Ia supernovae and
that from the early universe based on Planck satellite Cosmic Microwave Background 
observations (Riess, et al 2021).  Cepheids are also a benchmark for calculations of
evolutionary tracks, starting from their main sequence progenitors (B stars).   
B stars are more plentiful than O stars and the calculations do not have to include 
large amounts of mass loss.  Accurate 
modeling of  evolutionary tracks to the Cepheid stage is an important step in 
predicting and modeling exotic end stage objects resulting from a compact object 
 in a multiple system
(cataclysmic variables, some supernovae, and
 even gravitational waves).  Cepheids will typically become white dwarfs but the
 more massive ones may ultimately become  neutron stars.   Massive and intermediate-mass stars are frequently found in binary or multiple systems (e.g. Kervella, et al. 2019).    Specifically Cepheids have a binary/multiple fraction of greater than 57 $\pm$ 12 \%  (Evans, et al. 2022).

Ever since the first hydrodynamical pulsation calculations, it has been realized that 
masses predicted by evolutionary calculations are 10 to 20\% larger than those from 
pulsation calculations (see Neilson, et al. [2011] for a summary). The difference has 
been  somewhat reduced from investigations into opacities (e.g. Iglesias 
and Rogers 1991), core convective overshoot on
the main sequence (e.g. Prada Moroni et al. 2012), 
pulsation driven mass loss (Neilson, et al. 2011), and rotation (Anderson, et al. 2014).    
The effect of these parameters is complex.  For example, while increased 
convective overshoot produces an appropriate luminosity, it limits the 
temperature increase of the blue loops where the Cepheids are found so 
that they do not enter the instability strip for many masses.
It is likely that it is a combination of these which will reconcile the discrepancy.

Observational measurements of Cepheid masses are  needed to test these predictions.  
Since there are no Cepheids in eclipsing binaries in the Milky Way, 
there are two approaches available to measure masses.  First, 
interferometry has created a group of resolved binaries, from which
we can obtain the inclination i leading to both masses and distances (e.g. Gallenne, et al.
2019). The first result is a mass of 4.29 $\pm$ 0.13  M$_\odot$ for the Cepheid V1334 Cyg (Gallenne et
al, 2018), which uses velocities from the {\it Hubble Space Telescope (HST)} 
Space Telescope Imaging Spectrograph (STIS) high resolution spectra. It is significantly smaller
than predictions of evolutionary tracks (Evans, et al. 2018a).  Second, 
satellite ultraviolet (UV) spectroscopy with {\it HST} has provided a group of double-lined
spectroscopic binaries,  for example V350 Sgr (Evans, et al. 2018b). Using the orbital velocity amplitude of the Cepheid (from the ground) and the amplitude of a hot companion (from the UV) and combining these two with a mass for the main sequence companion inferred from the temperature provides the mass of the Cepheid.

A further motivation for determining the masses of Milky Way (MW) Cepheids as
accurately as possible is that 6 Cepheids (in 5 binary systems) 
have been found in eclipsing binaries in the Large
Magellanic Cloud (LMC; Pilecki, et al. 2018, 2021).  This means comparison between the
Mass--Luminosity (ML) relations for the metallicities of the MW and the LMC can now be made.
Even this high accuracy sample contains a warning.  Of the 5 systems, two may have  
 an evolutionary state different from the typical second or third crossing
of the instability strip:  first crossing and possible mass exchange systems.

This paper is a step toward  an additional method to determine Cepheid masses.
The mass function from a spectroscopic orbit  combines the orbital
period P and the  foreshortened semi-major axis of the Cepheid (A$_1$ sin i)
into a function involving the masses of the Cepheid M and the companion m,
and the inclination.  
For a reasonably
massive companion (a B star), the
companion dominates the spectrum in the far ultraviolet (FUV). Furthermore,
in this temperature range, the energy
distribution is very sensitive to temperature. This means the temperature
can be accurately determined, hence a mass
can be inferred, which is the topic of this study.

FN Vel is a classical Cepheid with a  pulsation period of 5.32$^d$ and a mean V magnitude of
10.3 mag.  It was discovered to be a spectroscopic binary by Anderson (2013) with an
orbital period of 471.654 $^d$ and a velocity amplitude of 21.899 km~s$^{-1}$.
The Ca II H and K lines were discussed by Kovtyukh et al. (2015).  From the relative
strength of the H and K lines, they showed that the companion is a hot star.

For the FN Vel system, an orbit has recently been provided (Shetye, et al. 2024).
Proper motions from Gaia can add the inclination (Kervella
et al 2019). However,  for the FN Vel system the orbital period of 1.29 years is shorter
than the observing windows for {\it Hipparcos} and {\it Gaia} DR3.  This means
the orbital velocity is smeared over the observing windows and is a lower limit
to the true orbital velocity. It is not until 
 the {\it Gaia} DR4 release when orbital motion will be
included in the solutions that the inclination will be reliable and tightly constrained.
At that time the combination of the mass function from the spectroscopic orbit, the mass
of the companion, and the inclination from {\it Gaia} will provide the mass for
the Cepheid itself.  

This paper is a discussion of an {\it HST} STIS spectrum of the Cepheid FN Vel in the 
1150 to 1700 \AA\/ region.     It is the first in a project to 
obtain FUV spectra of the 9 Cepheids with orbits (or at least orbital motion).
For  companions with B and A spectral types,
temperatures and masses will be determined.  For 
cooler companions an upper limit to the temperature will result, which will
add to the knowledge of the distribution of mass ratios in Cepheid systems.
 Masses will be derived from a study of masses of Detached Eclipsing Binaries (DEBs) in the same spectral region (Evans, et al. 2023)

The sections below discuss an {\it HST} STIS spectrum of the system, the energy
distribution, temperature and mass of the companion, and the orbit and limits on
the mass ratio and inclination of the system.

\section{Observation}

The FN Vel system was observed with the {\it HST} STIS Spectrograph on
December 23, 2019 for 2480 sec.
 The observation was with the
G140L grating ($R \approx 1000$)  with a wavelength range 1150 to 1700 \AA\/  in time-tag mode.


 This program was designed to observe binary Cepheids where the companion had not yet been detected, and where the exact separation of the companion from the Cepheid was uncertain. We therefore took a number of steps to improve the faint detection limit. These included observing at the D1 aperture position in the 52X2 aperture, which puts the source below the region of the FUV MAMA detector which can exhibit enhanced dark current, and observing in time-tag mode, which records the detection time of each photon. {\it HST} sees considerably reduced geo-coronal background when in Earth's shadow, and by using time-tag mode, we could potentially screen out the parts of the observation with the highest sky backgrounds. FN Vel turned out to be one of the brighter Cepheid companions found in this program, and so for this target these extra precautions were not needed.

 Data processing was done with the standard {\it HST} CALSTIS pipeline as described in Sohn et al (2019). As part of the extraction process for point sources, CALSTIS does a cross-correlation of the 2D flat-fielded spectral image using the expected shape of the spectral trace on the detector to centroid the spectrum and locate the extraction region to be used. Line lamp exposures are also taken to measure flexure in the STIS optical bench in both the dispersion and cross-dispersion direction. Since the narrow band F28X50OIII filter with an effective wavelength of about 5008 \AA\ was used for the target acquisition, it was the Cepheid itself rather than the hot companion that was placed at the 52X2D1 aperture location. However, comparing the results of the cross-correlation of the FUV spectrum with the expected spectral location as determined using the lamp images shows an offset of only about 3 mas, which is small compared to the expected uncertainty in the target acquisition. Thus the STIS data finds no significant offset between the Cepheid FN Vel and its hot companion, although this measure is only sensitive to offsets in the cross-dispersion direction, which in this case corresponds to a position angle of 266.15 degrees east of north.



\section{Energy Distribution}

\subsection{E(B-V)}








The reddening of the FN Vel system is significant, and important in interpreting the STIS spectrum.
The star is faint enough that it has not been as intensively studied as brighter Cepheids.  
Van Leeuwen, et al (2007) summarize the integrated mean photometry for the system $<$B$>$,   $<$V$>$, 
and  $<$I$_C$$>$. They list a reddening E(B-V) of 0.558 mag.  The photometry can also be used to calculate 
E(B-V) on the system of Dean, Warren, and Cousins (1978), using the relation from Fernie (1990).  This is 
useful since we know the photometry contains contributions from both the Cepheid and the hot companion.  Here
we explore whether the contribution from the companion affects the
derived reddening significantly.

Anticipating the result below,  the spectrum from 1100 to 1700 \AA\/
is well matched by the a temperature from the BOSZ Kurucz Atlas9 atmospheres
(Bohlin, et al.2017)  of 15000 K.
Using the calibration of Pecaut and Mamajek (2013) this is midway between B5
and B6 main sequence stars, which we will call B5.5
with B-V = -0.148 mag and (V-I)$_{C}$ = -0.155 mag. The calibration 
of Drilling and Landolt (2000; Table 15.7) provides M$_V$
of -1.1 mag for a B5.5 V star.
The absolute magnitude
of the Cepheid M$_V$  is -3.68 mag is from  Cruz Reyes and Anderson
(2023; Table 11) from
{\it Gaia} DR3 data for Cepheids
and open clusters.  Using the magnitude difference between the Cepheid and the B5.5 V 
star the correction to  $<$B$>$,   $<$V$>$, and  $<$I$_C>$ were computed in
Table~\ref{corr}.

From these, the original $<$B$>$ - $<$V$>$ and  $<$V$>$ - $<$I$_C>$ of 1.185
and 1.461 mag become  1.27 and 1.50 mag.  Using the appropriate reddening formula from Fernie (1990),
E(B-V) is  0.59 mag (Table~\ref{corr.red}).  That is, it is little affected by
the companion.






\begin{deluxetable}{lrrr}
\tablecaption{Correction for Companion\label{corr}}
\tablewidth{0pt}
\tablehead{
 \colhead{} &  \colhead{$<$ B $> $}  & \colhead{$<$ V$>$}  & \colhead{$<$ I$>$}  \\
}
\startdata
Cepheid + Companion  &   11.48  &   10.29       & 8.83 \\  
Cepheid &                11.68 &   10.39   &  8.88 \\
Companion  &             13.42 &   12.97  &   12.34 \\ 
\enddata
\end{deluxetable}

\begin{deluxetable}{lrrr}
\tablecaption{Reddening\label{corr.red}}
\tablewidth{0pt}
\tablehead{
 \colhead{} &  \colhead{$<$ B $>$ -  $<$ V$>$  }  & \colhead{$<$ V$>$ -  $<$I$>$}  & \colhead{E(B-V)}  \\
 \colhead{} &  \colhead{ mag }  & \colhead{ mag }  & \colhead{ mag }  \\
}
\startdata
Corrected   &   1.27       &   1.50    &   0.59    \\ 
Original  &    1.19  &  1.46  &  0.61   \\  
van Leeuwen     &  ---     &   ---      &  0.558  \\    
\enddata
\end{deluxetable}






\section{Companion Temperature}
The temperature of the companion FN Vel B was determined by comparison
of the spectrum with BOSZ model atmospheres computed from
Kurucz Atlas9 code (Bohlin et al. 2017).  The details of the
approach are presented in Evans, et al. (2023).  The 
spectrum is compared  with atmospheres through a series of
temperatures in steps of 500 K. The BOSZ models selected have
solar abundance, surface gravity (log g) of 4.0, microturbulence
of 2 km s$^{-1}$, and instrumental broadening of 500 km s$^{-1}$,
selected to match the STIS spectrum of a main sequence star.   

 Recently a revised version of the BOSZ models has been made available (Meszaros, et al. 2024).  The revisions include using both MARCS and Atlas9 atmospheres, as well as updated opacities.  We have investigated the differences the new models make.  Because of the increased opacity in the ultraviolet, the solutions are driven to higher temperatures.   The MK spectral B5 V standard $\rho$ Aur was  also found to be hotter than the Pecaut and Mamajek temperature for B5 V of 15700 K.  Although the 1150 to 1270 \AA\/ region is clearly sensitive to temperature, the temperatures derived from the models themselves have some uncertainty.  However, in this study of the companion to the Cepheid FN Vel, we use the temperature only to identify DEB spectra of the same temperature (from the same 2017 models) and deduce the mass from the comparison.  Thus the resulting mass for FN Vel B should be accurate, and we will continue to use the  2017 BOSZ models.

As discussed in Evans, et al. (2024), there are a number of strong
features of Si II and C II between 1250 and 1350 \AA, which are
temperature sensitive.  However, we do not attempt to model them
directly in the low resolution spectrum.

\begin{figure}
\plotone{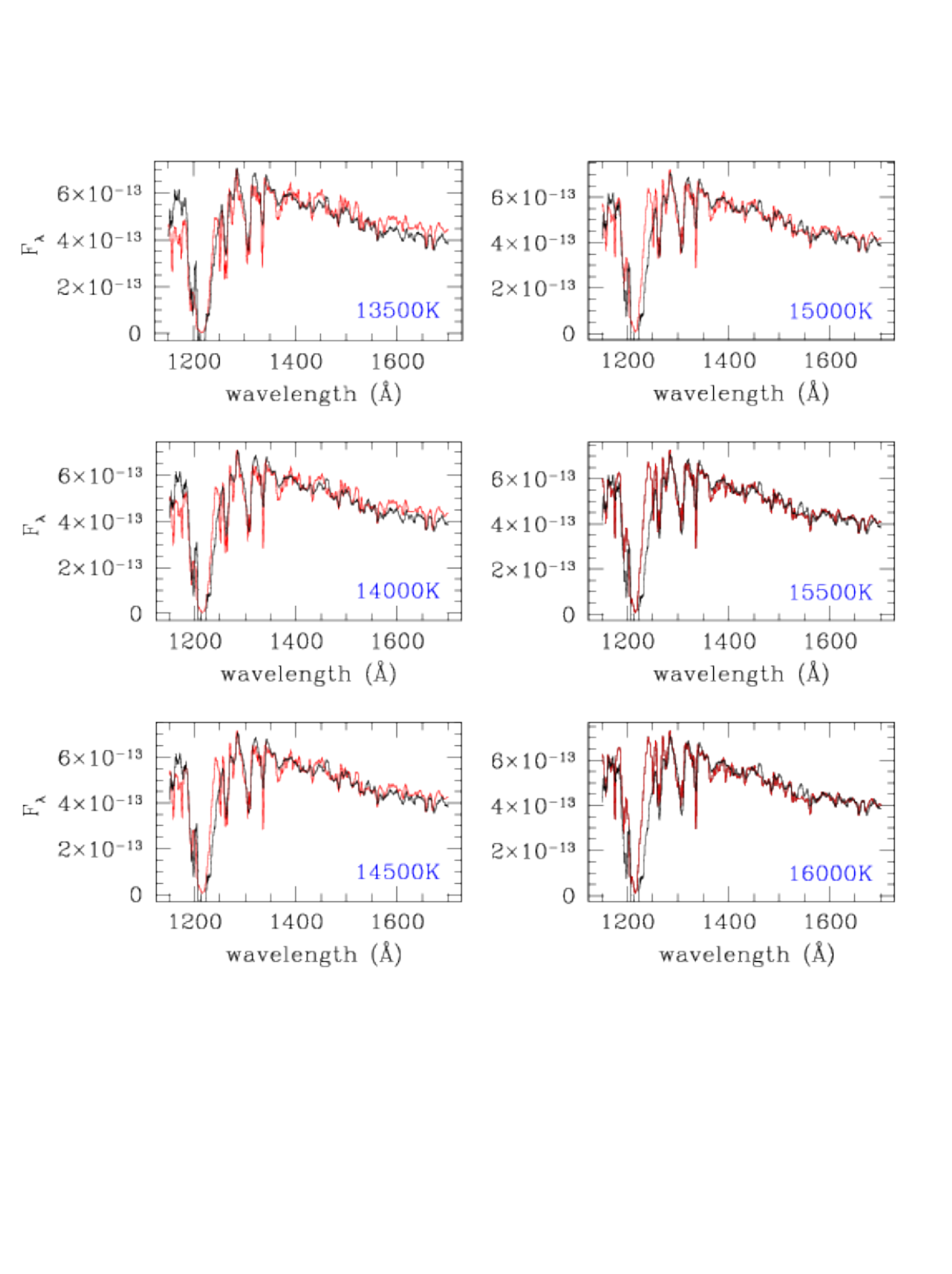}
\caption{Comparison between BOSZ models and the STIS  spectrum
  unreddened for E(B-V) = 0.60 mag.  Flux for all
  panels is in erg cm$^{-2}$ s$^{-1}$  \AA$^{-1}$; wavelength is in \AA.  Models are
  in red; spectrum is in black.  The temperature for the models is at the lower 
right of  each panel.  
\label{mod.comp}}
\end{figure}

\begin{figure}
\plotone{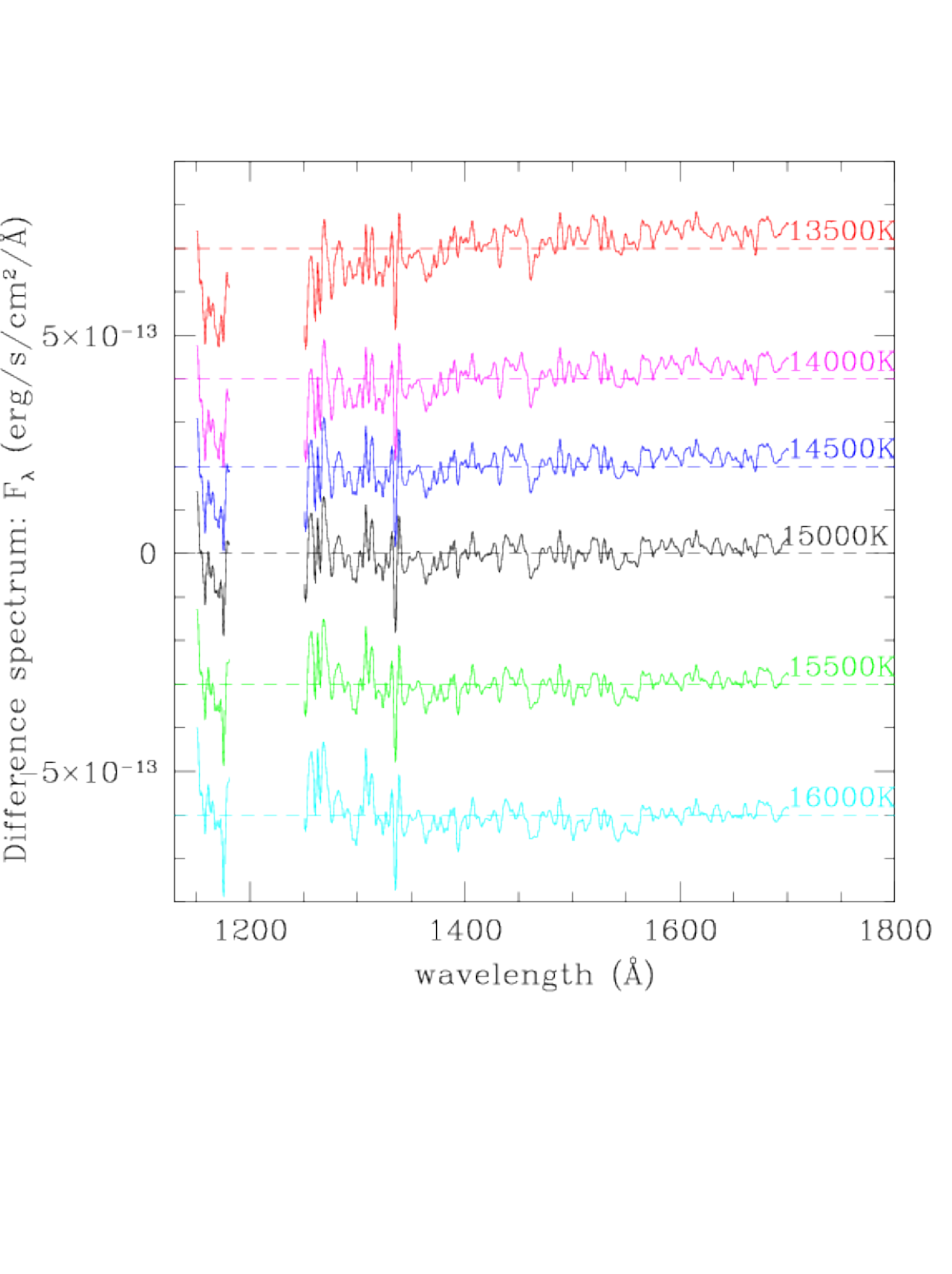}
\caption{The difference between the model and the spectrum for  E(B-V) = 0.60 mag.  The
  temperatures for the models are on the right of each difference spectrum.  
  The wavelengths between 1180 and 1250 \AA\/ are omitted because of
  contamination by interstellar Ly $\alpha$ absorption.  The difference spectrum in
  black is the best fit.  
\label{diff}}
\end{figure}

Fig~\ref{mod.comp} shows the spectrum (black) compared with models for
a series of temperatures.  Fig~\ref{diff} shows the spectrum differences
for these temperatures.   The Ly $\alpha$ region from
1180 to 1250 \AA\/ is omitted in the fit,
because of contamination from interstellar Ly $\alpha$.
  The progression
of the difference spectra from 13500 K to 16000 K illustrates the deficit
of flux at the shortest wavelengths in the models for the coolest models.
The hotter models have 
  increased  flux relative to the companion at the shortest
wavelengths.

\begin{figure}
\plottwo{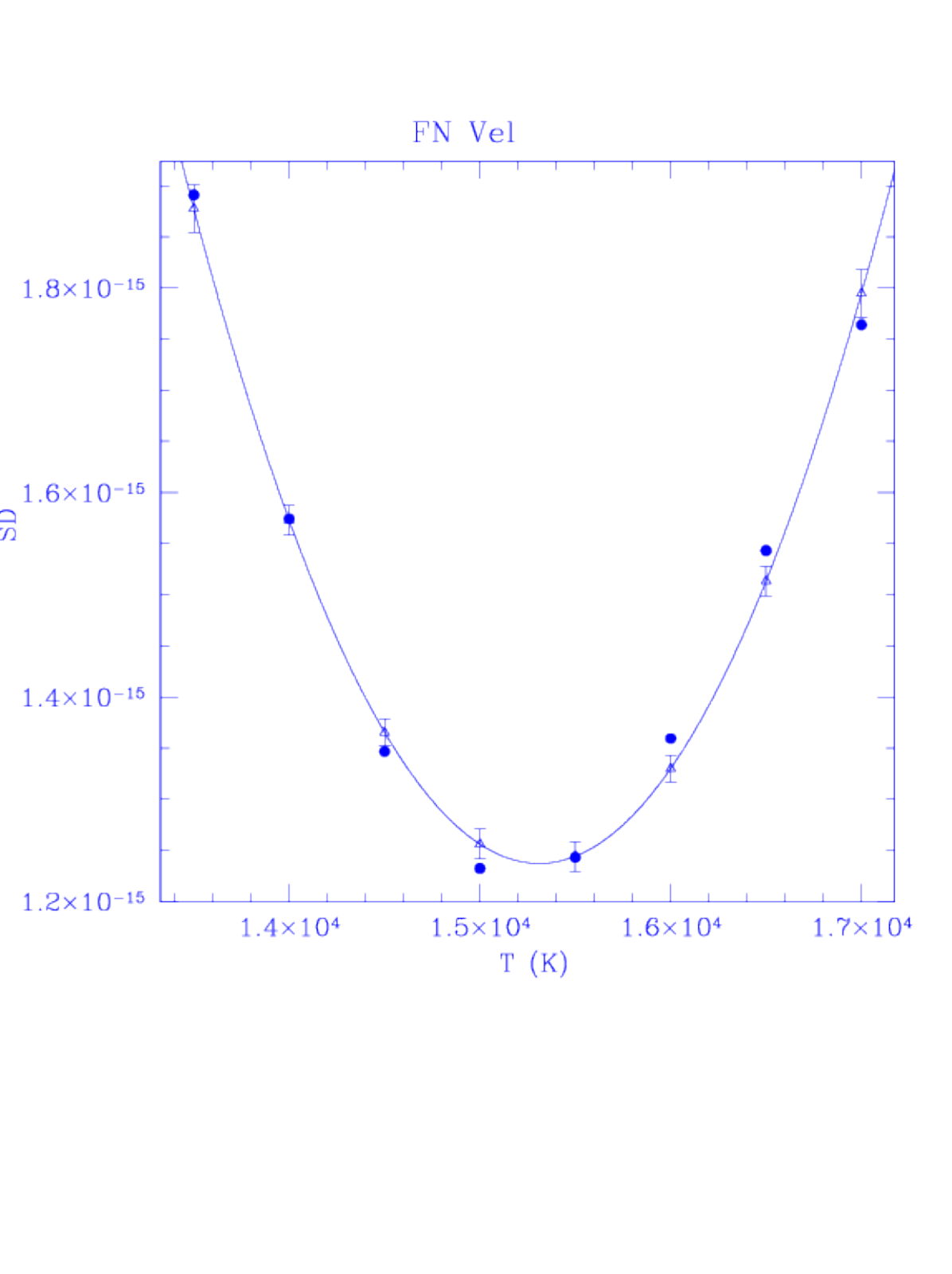}{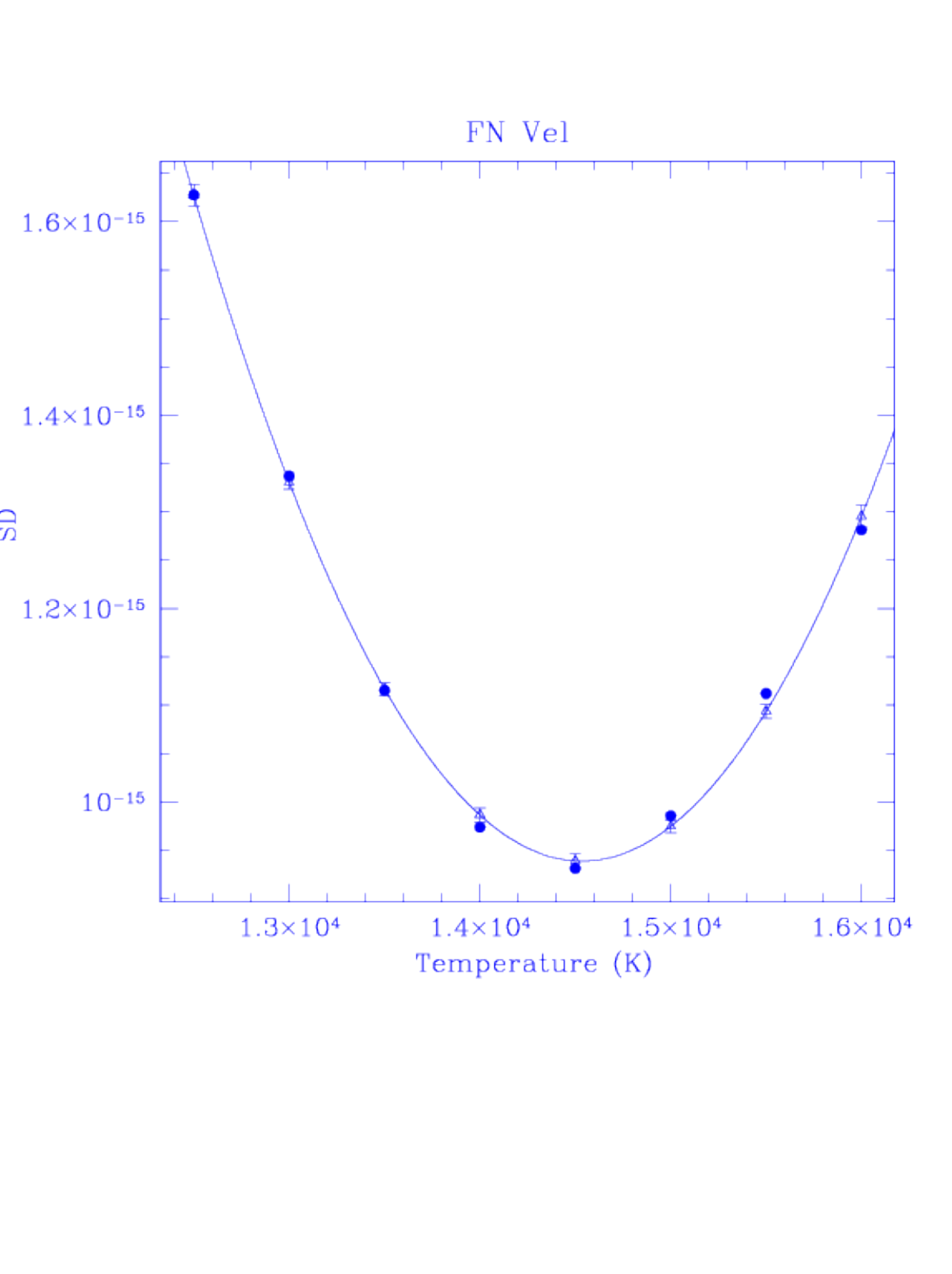}
\caption{  The standard deviations from the spectrum-model comparison
  as the temperature of the models is changed. Dots: the standard
  deviation; triangles: the parabola fit.  Left: E(B-V) = 0.60 mag.
  Right:  E(B-V) = 0.56 mag.
\label{par}}
\end{figure}

The determination of the best fit temperature is shown in Fig~\ref{par}
from the standard deviation of the difference spectra as a function of
temperature. The parabola fits are shown for two reddenings, E(B-V) = 0.60 mag
and 0.56 mag.  Table~\ref{tcom} lists the temperatures for these
reddenings.  The errors are the error from the parabola fit  (Err SD), and the errors
from the visual comparison of the difference spectra  (Err Vis), as discussed in
Evans, et al (2023). The difference in temperature for the two reddenings is
within the errors, confirming that the temperatures determinations are
not dominated by uncertainty in reddening.   Table~\ref{tcom} shows that
although the spectra are from the ultraviolet wavelength region, the energy
distribution is not highly sensitive to the reddening.  This is because the
spectrum only covers 600 \AA, and its shape is very sensitive to temperature,
particularly around Ly $\alpha$.  

\begin{deluxetable}{lrrr}
\tablecaption{FN Vel B Temperature\label{tcom}}
\tablewidth{0pt}
\tablehead{
 \colhead{E(B-V)} &  \colhead{T }  & \colhead{Err SD}  & \colhead{Err Vis}  \\
 \colhead{} &  \colhead{ K }  & \colhead{ K }  & \colhead{ K }  \\
}
\startdata
 0.56  &   14535  &   360  &   500  \\
 0.60  & 15391  &  629 &  500  \\
\enddata
\end{deluxetable}

\section{Companion Mass}

The relation between mass and temperature was determined for
Detached Eclipsing Binaries (DEBs) with  B and early A dwarfs (Evans, et al. 2023).
 In this study, {\it International Ultraviolet Explorer (IUE) satellite} spectra of DEBs in the same wavelength region as the STIS spectrum of FN Vel were compared with a series of BOSZ models to determine the temperature of the primary.  These temperatures were combined with the masses of the DEBs from Torres, et al. (2010) to derive the mass-temperature relation. 
The rms of the residuals from this relation is 0.037 in log M.  Rotation and abundance
difference contribute to the spread in this relation, as does age spread within
the main sequence.  An estimate
of the time on the main sequence (Ekstr\"om et al. 2012) results in  $\pm$ 0.04 in log M.  That
is, duration of the main sequence life could account for the entire spread. 
The companions of Cepheids are a special case within main sequence stars
since they are the companions of
more massive stars, and hence belong to the younger part of the DEB sample.  
For this reason, we use a Mass-Temperature relation to interpret Cepheids
which is 0.02 smaller in log M than that for DEBs.

Table~\ref{mfn} 
 illustrates the sensitivity of the mass of FN Vel B to the temperature from
 the STIS spectrum for a plausible range of temperatures.


\begin{deluxetable}{lr}
\tablecaption{Mass of FN Vel~B\label{mfn}}
\tablewidth{0pt}
\tablehead{
 \colhead{Temperature} &  \colhead{Mass }   \\
 \colhead{K} &  \colhead{  M$_\odot$ }   \\
}
\startdata
14900 &   3.89 \\
15400 &   4.09 \\
15900 &   4.31 \\
\enddata
\end{deluxetable}




\section{Orbit}

Further progress in obtaining the masses of both the Cepheid and the companion in the
FN Vel system makes use of a spectroscopic orbit.

\subsection{Spectroscopic Orbit}

\begin{figure}
\plotone{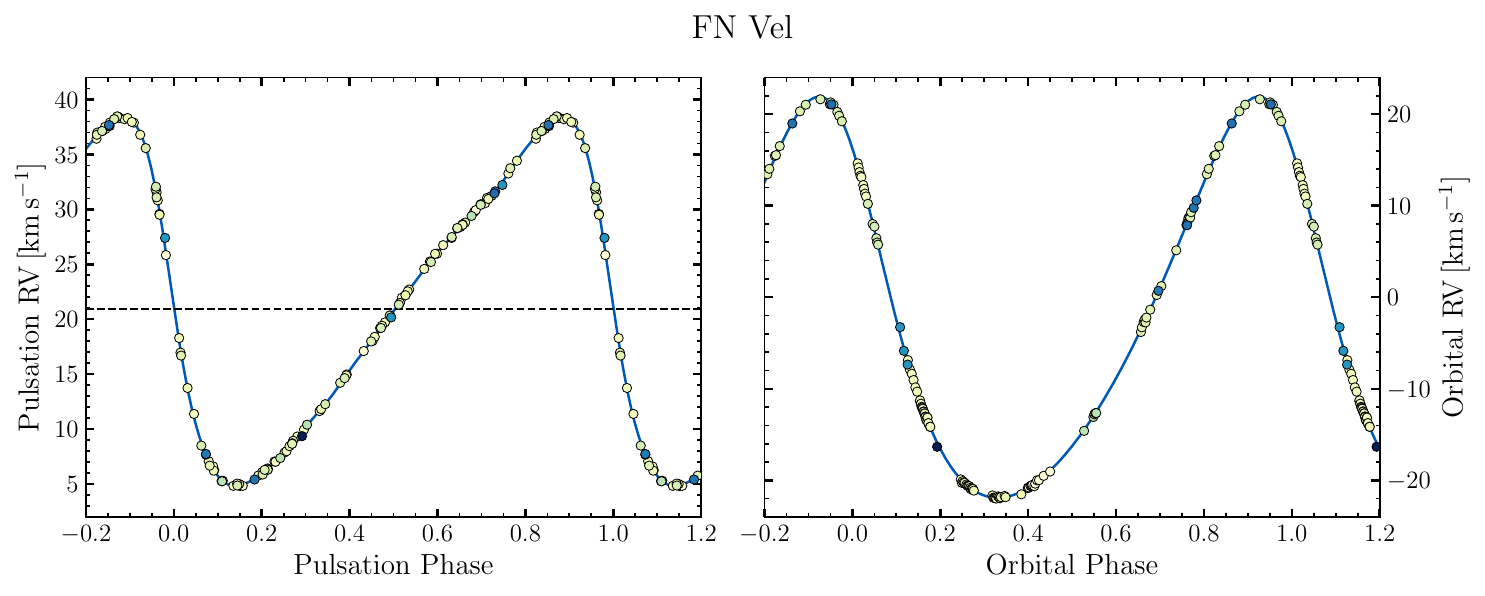}
\caption{ The pulsation velocity (left) and the orbital velocity (right)
  are shown as a function of their respective phases.  The dashed line in the left figure is the systemic velocity.)
The colors of the symbols indicate the date of observation   
from the earliest data in yellow to more recent data in blue.
(Adapted from Shetye, et al. 2024.)
\label{orb}}
\end{figure}

We discuss here briefly the spectroscopic orbit of the FN Vel system.
New radial velocities for the FN Vel binary since its discovery have been
obtained (Anderson, et al. 2024) in the VELOCE program.  An orbit has
been determined in Shetye, et al. (2024). The
results are shown in the Figure~\ref{orb} adapted from Shetye, et al.  The parameters from the
orbit are in Table~\ref{orb.param}.

The spectroscopic orbit provides the mass function:


$$F(M,m) = m^3 sin^3 i /(M + m)^2 = (A_1 sin i)^3/P^2$$ 


\noindent

where M and m are the masses of the Cepheid and the companion respectively
in solar masses,
P is the orbital period in years, 
and  (A$_1$ sin i) is the foreshortened semi-major axis of the Cepheid
in au.  

The remaining parameter needed to determine the Cepheid mass is the inclination.
Ultimately, this will be provided by {\it Gaia}.
Proper motions from {\it Gaia} and those from {\it Hipparcos}
(and a distance) combined with a spectroscopic orbit provide
complete velocity data for the orbit, and hence  
an inclination  as discussed by Kervella, et al. (2019).  As pointed out there,
the proper motions from those two catalogs are not instantaneous, but
an average over the observing window.  For Cepheids with orbital periods
shorter than those observing windows (as is the case for FN Vel), the smearing
results in a decreased proper motion.  The smearing can be imperfectly
approximated (Kervella, Arenou, Mignard, and Th\'evenin 2019) for the
ratio of the orbital period to the {\it Gaia} DR3 observing window (472/1038).
Unfortunately this estimate produces an inclination too uncertain to
provide usable results. 
This will be revisited when the
  {\it Gaia} DR4 release includes astrometric orbital fitting. Furthermore, the longer
  observation stream in the {\it Gaia}
DR4 release will remove the dependency on the relatively large
{\it Hipparcos} uncertainty.





\newcommand{\hhh}{\hspace*{0.070in}}
\newcommand{\iii}{\hspace*{0.015in}}

\begin{deluxetable}{lrr}
\tablecaption{FN Vel Orbit\label{orb.param}}
\tablewidth{0pt}
\tablehead{
}
\startdata
P$_{orb}$ &(yr)  &  1.2917 $\pm$ 0.0002   \\
T$_{0}$ & (JD)   & 2,456,407.87 $\pm$  0.36  \\
e &     & 0.218 $\pm$  0.001   \\
K & (km sec$^{-1}$)    & 21.91  $\pm$  0.02 \\
$\omega$ &  ($^0$)   & 42.07  $\pm$  0.21 \\
A$_1$ sin i  & (au)      &  0.927  $\pm$ 0.001  \\
f(m) &   (M$_\odot$) &  0.48  $\pm$ 0.0017 \\
\enddata
\end{deluxetable}

\section{Discussion and Conclusions}

In summary, in the FN Vel binary system the mass of the companion has been
determined from the energy distribution from a STIS ultraviolet spectrum.  An
orbit is now available.  The remaining parameter required to determine
the mass of the system is the inclination of the orbit. This will be
available from the {\it Gaia} DR4 data release.  

\begin{figure}
\plotone{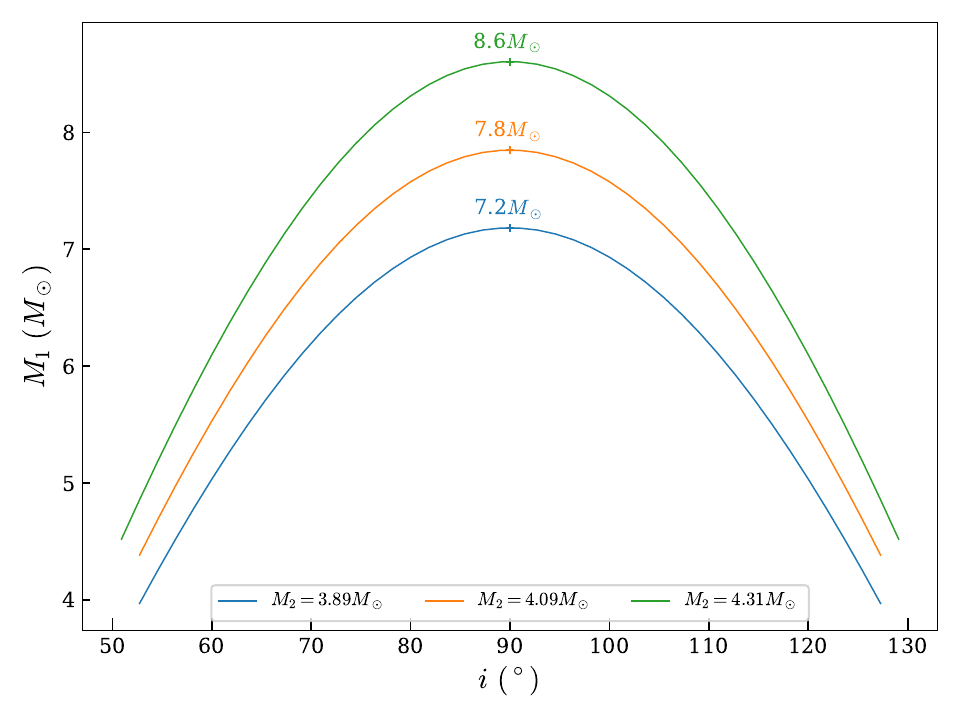}
\caption{The Cepheid mass as a function of inclination and the mass of the
  companion.  The lines show the relation
  for the range of companion masses M$_2$ in Table~\ref{mfn}: 3.89 M$_\odot$: blue;
  4.09 M$_\odot$: red;  4.31 M$_\odot$: green.  The maximum value of the Cepheid
  mass M$_1$ is listed above the curves for the companion masses M$_2$
\label{limits}}
\end{figure}

With the available data, some limits can be placed on the Cepheid mass.  Using the
mass function from Table~\ref{orb.param} with the companion mass from Table~\ref{mfn} 
the Cepheid mass can be calculated for values of the inclination.
Fig.~\ref{limits} shows that for the inclination  between 50 and 130 $^o$, the
Cepheid mass is between 4 and 9 M$_\odot$

To connect the Cepheid FN Vel with its past history, the  Achernar system appears
to be a good model for a progenitor
(Kervella et al. 2022).  It is a Be star (A) with an A star
companion (B).  The mass for A has recently been determined
to be 6.0 $\pm$ 0.6 M$_\odot$ with an orbital period of 7 years.
The separation between A and B is large enough that 
it is anticipated
that A will evolve to become a Cepheid without mass exchange between the components.
Because the orbital period of of FN Vel is shorter, the question is raised whether
 mass exchange within the system has occurred in the past,
particularly when the primary was at the tip of
the red giant branch.  Systems such as FN Vel are a good probe of this question.
One indication of past history is the summary of orbit parameters
for Cepheids (Shetye et al
2024).  The orbital period of FN Vel is just on the boundary of periods where Cepheid
systems exist.  With a pulsation period of 5 days, it is not an unusually large
Cepheid.  Furthermore the system has not been fully circularized.  Thus the system
seems to have  at most had  mild mass exchange.  However as an additional system
very close to the limit of mass exchange it is a useful data point. 

To summarize, after the {\it Gaia} DR4 release,
the combination of a mass for the secondary FN Vel B, a spectroscopic
orbit, and an inclination will provide a mass for the Cepheid, FN Vel A.  This is
valuable  for comparison with evolutonary calculations.



\section{Acknowledgments}

Support for JK and CP were provided from HST-GO-15861.001-A.
Support was provided to NRE by the Chandra X-ray Center NASA Contract NAS8-03060.
HMG was supported through grant HST-GO-15861.005-A from the STScI under NASA contract NAS5-26555.
This project has received funding from the European Research Council (ERC) under the European Union’s Horizon 2020 research and innovation programme  to PK (projects CepBin,
grant agreement No 695099, and UniverScale, grant agreement No 951549).
RIA is funded by the SNSF through an Eccellenza Professorial Fellowship, grant number PCEFP2\_194638.
This project has received funding from the European Research Council (ERC) under the European Union's Horizon 2020 research and innovation programme (Grant Agreement No. 947660). SS would like to acknowledge the Research Foundation-Flanders (grant number: 1239522N).
AG acknowledges the support of the Agencia Nacional de Investigaci\'on Científica y Desarrollo (ANID) through the FONDECYT Regular grant 1241073.
This work has made
use of data from the European Space Agency (ESA) mission Gaia (https://www.cosmos.esa.int/gaia), processed by
the Gaia Data Processing and Analysis Consortium (DPAC, https://www.cosmos.esa.int/web/gaia/dpac/consortium).
Funding for the DPAC has been provided by national institutions, in particular the institutions participating in the
Gaia Multilateral Agreement.
The SIMBAD database, and NASA’s Astrophysics Data System Bibliographic Services
were used in the preparation of this paper.

The data presented in this article were obtained from the Mikulski Archive for Space Telescopes (MAST) at the Space Telescope Science Institute. The specific observations analyzed can be accessed
via \dataset[10.17909/m493-xr06]{https://doi.org/DOI}






\begin{thebibliography}{}

\bibitem[\protect\citeauthoryear{anderson}{2013}]{and13}
  Anderson, R. I. 2013, PhD Thesis, Univ of Geneva
  
\bibitem[\protect\citeauthoryear{anderson etal}{2014}]{and14}
Anderson,. R. I., Ekstr\"om, S., Georgy, C. et al. 2014, A\&A 564, A100 


\bibitem[\protect\citeauthoryear{anderson etal}{2023}]{and24}
Anderson, et al. 2024, A\&A, 686, A177

\bibitem[\protect\citeauthoryear{bohlin etal}{2017}]{bol17}
Bohlin, R. C., Meszaros, S. Fleming, S. W., Gordon, K. D., Koekemoer, A. M., and Kovacs, J. 2017, A. J.,
153, 234


\bibitem[\protect\citeauthoryear{breuval etal}{2020}]{bre20}
Breuval, L., Riess, A. G., Casertano, S., et al. 2024, arXiv.2404.08038



\bibitem[\protect\citeauthoryear{cruz reyes anderson}{2023}]{cra23}
Cruz Reyes, M. and Anderson, R. I. 2023, A\&A, 672, A85


\bibitem[\protect\citeauthoryear{dean etal}{1978}]{dwc78}
Dean, J. F., Warren, P. R., and Cousins, A. W. J. 1978, MNRAS, 183, 569

\bibitem[\protect\citeauthoryear{drilling landolt}{2000}]{dri00}
Drilling, J. S., and Landolt, A. U. 2000, in Astrophysical Quantities, ed.A. N. Cox(New York: Springer), 381

\bibitem[\protect\citeauthoryear{ekstroem etal}{2012}]{eks12}
Ekstr\"om. S., Georgy, C, Eggenberger, P., et al. 2012, A\&A, 537, A146


\bibitem[\protect\citeauthoryear{evans etal}{2018}]{ev18a}
  Evans, N. R., Karovska, M., Bond, H. E., et al. 2018a, ApJ, 863, 187

 \bibitem[\protect\citeauthoryear{evans etal}{2018}]{ev18b} 
  Evans, N. R., Proffitt, C., Carpenter, K. G. et al. 2018b, ApJ, 866, 30


\bibitem[\protect\citeauthoryear{evans etal}{2023}]{ev23}
Evans, N. R., Ferrari, M. G., Kuraszkiewicz, J. et al 2023, AJ, 166, 109  

\bibitem[\protect\citeauthoryear{evans etal}{2024}]{ev24}
  Evans, N. R., Gallenne, A., Kervella, P. et al. 2024, ApJ, 972, 145

\bibitem[\protect\citeauthoryear{evans etal}{2022}]{ev22}
  Evans, N. R., Engle, S., Pillitteri, I., et al. 2022, ApJ, 938, 153

\bibitem[\protect\citeauthoryear{fernie}{1990}]{fer90}
Fernie, J. D.  1990, ApJS, 72, 153

\bibitem[\protect\citeauthoryear{gallenne etal}{2019}]{gal19}
Gallenne, A., Kervella, P., Borgniet, S.  et al. 2019, A\&A, 622, 164

\bibitem[\protect\citeauthoryear{gallenne etal}{2018}]{gal18}
Gallenne, A., Kervella, P., Evans, N. R.,  et al. 2018, ApJ, 867, 121

\bibitem[\protect\citeauthoryear{iglesias rogers}{1996}]{igl96}
Iglesias, C. A., and Rogers, F. J. 1996, ApJ, 464, 943

\bibitem[\protect\citeauthoryear{kervella et al}{2019}]{ker19}
Kervella, P., Gallenne, A. Evans, N. R. et al. 2019, A\&A, 623, A116

\bibitem[\protect\citeauthoryear{kervella arenou mignard thevenin}{2019}]{kamt19}
Kervella, P., Arenou, F., Mignard, F., and Th\'evenin, F. 2019, A\&A, 623, A72

\bibitem[\protect\citeauthoryear{kervella et al}{2022}]{ker22}
Kervella, P., Borgniet, S., Domiciano de Souza, A. et al. 2022, A\&A, 667, A111

\bibitem[\protect\citeauthoryear{kovtyukh etal}{2015}]{kov15}
Kovtyukh, V. Szabados, L., Chekhonadskikh, F., Lemasle, B., and Belik, S.
2015, MNRAS, 448, 3567




\bibitem[\protect\citeauthoryear{meszaros etal}{2024}]{mes24}
Meszaros, S., Bohlin, R., Prieto, C. A. et al. 2024 A\&A, 688, 197

\bibitem[\protect\citeauthoryear{neilson etal}{2011}]{nei11}
Neilson, H. R., Cantiello, M., and Langer, N. 2011, A\&A 529, L9

\bibitem[\protect\citeauthoryear{pecaut mamajek}{2013}]{pec13}
Pecaut, M. J., and Mamajek, E. E. 2013, ApJS, 208, 9

\bibitem[\protect\citeauthoryear{pilecki etal}{2018}]{pil18}
 Pilecki, B., Gieren, W.,  Pietrzynski, G., et al. 2018, ApJ, 862, 43


 \bibitem[\protect\citeauthoryear{pilecki etal}{2021}]{pil21}
 Pilecki, B., Pietrzynski, G., Anderson, R. I., et al. 2021, ApJ, 910, 118

 \bibitem[\protect\citeauthoryear{prada moroni etal}{2012}]{pra12}
Prada Moroni, P. G., Gennaro, M., Bono, G., et al. 2012, ApJ, 749, 108

\bibitem[\protect\citeauthoryear{riess et al}{2022}]{rie22}
Riess, A. G. et al. 2022 , ApJ, 934, L7



\bibitem[\protect\citeauthoryear{shetye et al}{2024}]{she24}
Shetye, S., Anderson, R. I., Viviani, G. et al 2024, arXiv:2405.19840

 \bibitem[\protect\citeauthoryear{sohn et al}{2019}]{soh19}
Sohn, S. T., et al., 2019, “STIS Data Handbook”, Version 7.0, (Baltimore: STScI)

\bibitem[\protect\citeauthoryear{torres}{2010}]{tor10}
Torres, G., Andersen, J., and Gimenez, A. 2010,A\&ARv,18, 67

\bibitem[\protect\citeauthoryear{vanleeuwen etal}{2007}]{vanl07}
Van Leeuwen, F. Feast, M. W., Whitelock, P. A., and Laney, C. D. 2007, MNRAS, 379, 723 


\end{thebibliography}
\end{document}